\documentclass{pnastwo}

\url{www.pnas.org/cgi/doi/10.1073/pnas.0709640104}
\copyrightyear{2008}
\issuedate{Issue Date}
\volume{Volume}
\issuenumber{Issue Number}
\usepackage{bm}

\newcommand\red[1]{{\color{black}#1}}

\begin{document}

\title{A Topologically Driven Glass in Ring Polymers}

\author{Davide Michieletto\affil{1}{School of Physics and Astronomy, University of Edinburgh, Edinburgh, United Kingdom} and
Matthew S. Turner\affil{2}{Department of Physics, University of Warwick, Coventry, United Kingdom}}

\contributor{Submitted to Proceedings of the National Academy of Sciences
of the United States of America}

\significancetext{The glass transition is commonly associated with a reduction in the temperature of liquids or by an increase in density of granular materials. In this work we propose a radically different approach to study dynamical arrest that relies on the topology of the components. We find that a concentrated solution of ring polymers can be driven to a kinetically arrested state by randomly pinning a small fraction of rings, a transition not observed in linear polymers. We attribute this jamming to topological interactions, called ``threadings'', that populate solutions of rings. Our work provides the first evidence for these threadings and suggests that very long rings may be expected to be kinetically arrested even as the fraction of pinned rings approaches zero.}

\maketitle

\begin{article}
\begin{abstract}
{The static and dynamic properties of ring polymers in concentrated solutions remains one of the last deep unsolved questions in Polymer Physics. At the same time, the nature of the glass transition in polymeric systems is also not well understood. In this work we study a novel glass transition in systems made of circular polymers by exploiting the topological constraints that are conjectured to populate concentrated solutions of rings. We show that such rings strongly inter-penetrate through one another, generating an extensive network of topological interactions that dramatically affects their dynamics. We show that a kinetically arrested state can be induced by randomly pinning a small fraction of the rings. This occurs well above the classical glass transition temperature at which microscopic mobility is lost. Our work demonstrates both the existence of long-lived inter-ring penetrations and also realises a novel, topologically-induced, glass transition.}
\end{abstract}


\dropcap{T}he physics of ring polymers remains one of the last big mysteries in Polymer Physics~\cite{McLeish2008}. 
Concentrated systems of ring polymers have been observed in both, simulations and experiments, to display unique features that are not easily reconciled with the standard reptation theory of linear polymers~\cite{Gennes1979,Doi1988,Halverson2011,Halverson2011a,Kapnistos2008}. The main reason for this is that ring polymers do not possess free terminal segments, or ends, essential for end-directed curvilinear diffusion. In contrast, ring polymers possess a closed contour, which leads to a markedly different relaxation and diffusion mechanisms. Recently, there has been much improvement in the production of purified systems of rings~\cite{Kapnistos2008,Pasquino2013,Doi2015} with the consequent result that more and more experimental puzzling evidence requires a deeper understanding of their motion in concentrated solutions and melts from a theoretical point of view. 

Recently, it has been conjectured that ring polymers assume crumpled, segregated conformations in concentrated solution or the melt~\cite{Halverson2011a}. 
On the other hand, numerical and experimental findings~\cite{Halverson2011a,Kapnistos2008}, suggest that rings exhibit strong inter-coil correlations which have proved difficult to address in simplified theoretical models~\cite{Cates1986,Rubinstein1986,Grosberg2014,Smrek2015}. Because of this, there have been many recent attempts to rigorously characterise these inter-chains interactions~\cite{Michieletto2014,Michieletto2014a,Tsalikis2014,Lee2015}, although a precise definition and unambiguous identification of these ``threadings'' in concentrated solutions of rings remains elusive. The primary reason for this is that the rings are assumed to remain strictly topologically un-linked from one-another throughout if synthesised in this state.

In the case of concentrated solutions of rings embedded in a gel, a method to identify these inter-penetrating threadings has recently been proposed~\cite{Michieletto2014}. Here it was shown that the number of threadings scales extensively in the polymer length (or mass) and can therefore be numerous for long rings, creating a hierarchical sequence of constraints that can span the entire system.
It has also been conjectured that a kinetically frozen state, or a ``topological glass''~\cite{Lo2013} can emerge, since such an extensive network of constraints can eventually suppress the translational degrees of freedom of the rings. However the molten, or highly concentrated, state does differ from that of polymers embedded in a gel and so whether a similar jamming transition occurs for long enough polymers or even whether threadings are present in the absence of a gel remain open problems and are the main questions addressed in this study. An example of inter-threaded ring configuration is shown in Fig.~1. A spherical region (Fig.~1(b)) is carved from the configuration depicted in Fig.~1(a) which represents a typical system studied in this work. The degree of inter-penetration between the different coils is readily appreciable from the figure, and it can be boiled down even further into a network representation~\cite{Michieletto2014}.  
The un-crossability constraint between chains transforms the threadings into topological hindrance in the motion of the coils, which we conjecture to form the basis for a dramatic slowing down in the dynamics of long enough coils.

Glass-forming systems exhibit degrees of freedom that become constrained as the temperature, or the density, of the system approaches the glass transition temperature $T_g$ or the critical jamming density $\rho_c$, thereby (super-)exponentially increasing the viscosity of the system~\cite{Gibbs1965}. Understanding the origin of these constraints, being kinetic or thermodynamic in nature, is still an open topic that animates intense research~\cite{Berthier2011,Ritort2003}.

Recently, a novel and promising theoretical approach to study the glass transition in glass-forming liquids has been advanced: It involves perturbing a system by randomly pinning some fraction of the constituents and by observing the behaviour of the un-frozen fraction. This method introduces a field of ``quenched'' disorder by freezing in space and time a subset of the system ~\cite{Biroli2008,Cammarota2012,Karmakar2013,Gokhale2014,Nagamanasa2014,Ozawa2015,Kob2014}.

Inspired by this approach we focus our attention on a concentrated solution of rings and apply a similar protocol: we freeze in space and time a fraction of polymers in the system and observe the response of the un-frozen constituents.  We find that while linear polymers are substantially insensitive to this perturbation, ring polymers become irreversibly trapped in a network of inter-coil constraints (threadings) which, in the limit of long rings, allows us to drive a kinetically arrested state with only a small fraction of permanently frozen chains. We conjecture that a spontaneous glassy state might therefore emerge in the long chain limit, even as the fraction of explicitly pinned chains goes to zero.  Because these constraints are topological in nature, originating from non-crossability of the chains, this glassy state has the potential to be produced at arbitrary temperature or monomer density, provided only that these inter-coil topological interactions remain abundant in the system, \emph{i.e.}, the rings are sufficiently long and not too dilute. This system is therefore a candidate for a novel kind of glass transition in systems made of polymers or other elements with non-trivial topology. 

\begin{figure*}[t]
\centering
\includegraphics[width=1.0\textwidth]{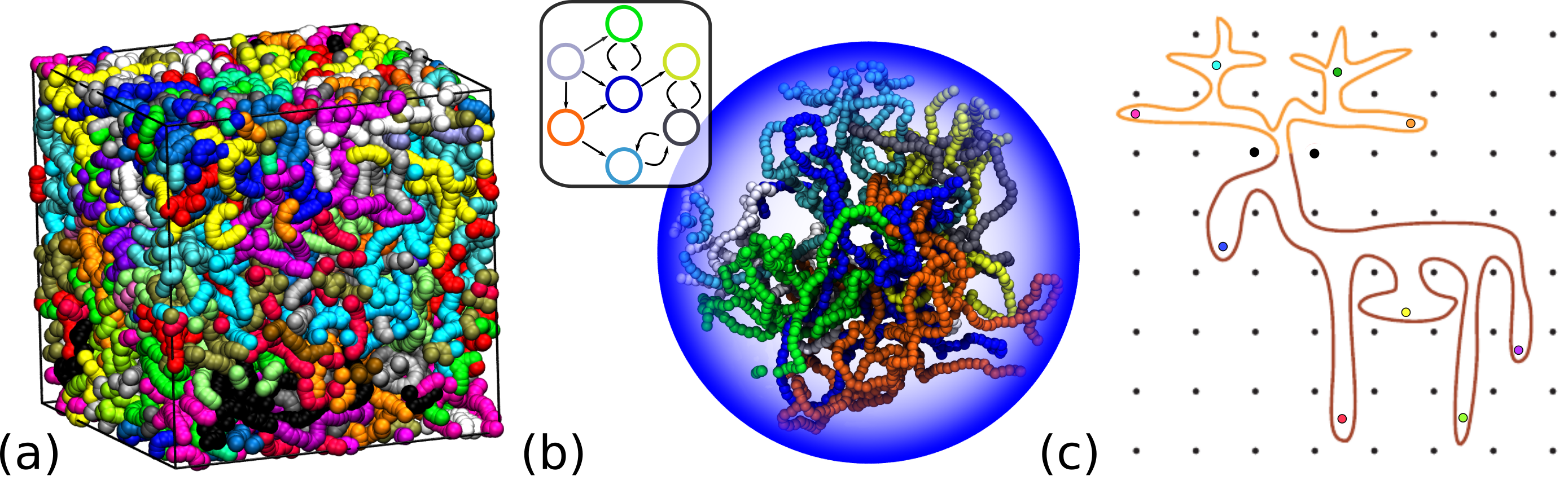}
\caption{ {\bf Networks of threadings} \textbf{(a)} Snapshot from a simulation representing a system with $N=50$ chains $M=512$ beads long. \textbf{(b)} Snapshot of a spherical region and some of the rings from \textbf{(a)}. One can easily appreciate the inter-penetration between the rings in the system (for instance cyan through dark grey or orange through blue). The sketched network in the top corner is obtained by visual inspection and depicts the penetrations of the rings. Two circles are connected by a directed arrow when the former is threading through the latter~\cite{Michieletto2014} (see also Suppl. Movie M1). %Two rings can mutually thread. 
%\textbf{(b)} Idealised configuration of rings, where levels of hierarchical threadings are enumerated starting from the innermost (black) ring. This structure requires a high level of cooperativity in order to be broken, as discussed in the text.
% rings of hierarchy $n+1$ must be moved before those of hierarchy $n$ are un-pinned; 
(c) Lattice animal representation of a single ring polymer surrounded by non-threading rings (black dots) and penetrated by threading rings (coloured dots). Modified from Ref.~\cite{Kapnistos2008} with permission.
}
\label{fig:Panel_Snap_Threading}
\end{figure*}

\vspace*{-0.5 cm}
\subsection{Rings in Solution Assume Crumpled but Largely Overlapping Conformations}
The static properties of rings in solution have been studied for some decades and it is nowadays  thought that rings assume crumpled conformations characterised by a scaling law for their gyration radius $R_g \sim M^\nu \text{ with }\nu \simeq 1/3
\label{eq:Rg_scaling}$
in the limit of rings with large polymerisation $M$~\cite{Grosberg2014,Grosberg1993,Halverson2014,Bras2014,Rosa2014}. On the other hand they also have been found to maintain numerous contacts with their neighbours. This is often quantified by measuring the ``contact surface'' of the polymers, which is defined as the number of segments $m_{\rm surf}$ in any one chain that are in contact with any segment belonging to any other chain. This has been found to scale nearly extensively with the size of rings (SI Appendix, Figs.~S1-S2) %\ref{fig:SI_SurfaceMonomers}  and Ref.~\cite{Halverson2011,Halverson2014}), \emph{ i.e.} 
$m_{\rm surf} \sim M^{\beta} \text{ with }\beta\simeq 0.98
\label{eq:m_surf_scaling}$.
This has been interpreted as a clear signature of abundant interactions and coil overlap. This picture is also supported by the fact that a coil's pair correlation function $g(r)$ is peaked at $r_c < 2 r_{\rm max} \simeq 2.6 R_g$ (see Fig.~S3 %~\ref{fig:SI_Gr} 
in SI), in light of which the coils can be viewed as inter-penetrating ultra-soft colloids of radius $r_{\rm max} \simeq 1.3 R_g$~\cite{Bernabei2013,Likos2014,Mattsson2009}. 

The counter-intuitive fact that rings have numerous inter-coil interactions while assuming a scaling exponent $\nu\simeq 1/3$ can be understood within the fractal globule conjecture~\cite{Mirny2011,Grosberg2014} in which the rings assume a fractal, hierarchical and virtually entanglement-free, conformation, that can accommodate near extensive interactions with other chains with a contact exponent $\gamma$ near unity~\cite{Halverson2014} (SI Appendix Fig.~S3). 

These inter-chain interactions may include ``threadings'' that have been conjectured to be intimately related to the slow overall diffusion of the rings centre of mass~\cite{Halverson2011a,Michieletto2014,Lee2015,Doi2015} (see also SI Appendix Fig.~S4), an observation that is in apparent contrast with the very fast stress relaxation~\cite{Kapnistos2008,Pasquino2013}, characterised by a power-law decay of the relaxation modulus $G(t)$ and by a remarkable absence of the entanglement plateau that characterises concentrated solutions of linear polymers.
In light of these findings, we believe that further investigation of the role of these inter-ring interactions is crucial.

\subsection{Contiguity is Persistent for Longer Chains} \mbox{In order} to probe inter-coil interactions we first adopt a definition of surface monomers where the $i$-th monomer of chain $I$ is a surface monomer of that chain if its distance from a monomer $j$ belonging to a different chain $J$ is $d_{ij}< \rho^{-1/3}$ where $\rho=0.1 \sigma^{-3}$ is the monomer concentration. We then define as ``contiguous'' two coils that share surface monomers. 
From this definition, we propose a method to track the exchange dynamics of contiguous chains, $I$ and $J$, by computing a dynamic $N\times N$ matrix $P(t)$ whose elements are defined as 
\begin{equation}
P_{IJ}(t) = \begin{cases}
0 \text{ if } d_{ij} \geq \rho^{-1/3} \quad \forall~\text{ }i, j\\
1 \text{ otherwise} 
\end{cases}
\label{eq:ContactMatrix}
\end{equation}
From this it is straightforward to obtain the correlation function
\begin{equation}
\varphi_{nc}(t) = \left\langle \dfrac{1}{N}\sum_{J=1}^N P_{IJ}(t)P_{IJ}(t-\Delta t)\dots P_{IJ}(0) \right\rangle
\label{eq:PersNeigh}
\end{equation}
where $\langle \dots \rangle$ indicates the ensemble average over rings $I$ and initial times. This function quantifies the exchange dynamics of contiguous chains and tracks the time that the chains {\em first} become non-contiguous since it involves the product of $P_{IJ}$ over all the intermediate time-steps up to time $t$. The behaviour of $\varphi_{\rm nc}(t)$ is reported in Fig.~2.
Expression \eqref{eq:PersNeigh} gives a more {strict} measurement of the inter-chain cooperativity than would be obtained from a standard contiguity correlation function and, unlike the latter, should decay to zero over time-scales that are comparable with the time taken for the chains centre of mass to diffuse away from one-another. 
While for short chains this is well described by a simple exponential, we observe that its behaviour for longer chains can be fitted for \red{about two} decades by stretched exponentials 
\begin{equation}
\varphi_{\rm nc} = \exp{\left[-\left(\dfrac{t}{\tau_{\rm nc}}\right)^{\beta_{\rm nc}}\right]}
\label{eq:Stretchexp}
\end{equation}
with an exponent $\beta_{\rm nc}$ which varies from $\beta_{\rm nc}=1$ for $M=256$ to $\beta_{\rm nc}\simeq 1/2$ for the longest chains in the system (see Fig.~S5 in SI). Even more striking is the exponential increase of the typical time to become non-contiguous $\tau_{\rm nc}$, indicating a very slow de-correlation between chains (or very long ``exchange time''), often interpreted as the onset of glassy dynamics~\cite{Mattsson2009}. As $\varphi_{\rm nc}$ shows fat tails at long times, we also compute $T_{\rm nc}$ as the (numerical) integral of the correlation function. We find that the functional behaviour of $T_{\rm nc}$ is in agreement with $\tau_{\rm nc}$ and shows an even steeper increase (see inset Fig.~2). This is most likely due to the fact that only the long-time tails show deviations from the stretched exponential behaviour, and these act to further increase the exchange time of the coils. 

\begin{figure}[h]
\centering
\includegraphics[width=0.5\textwidth]{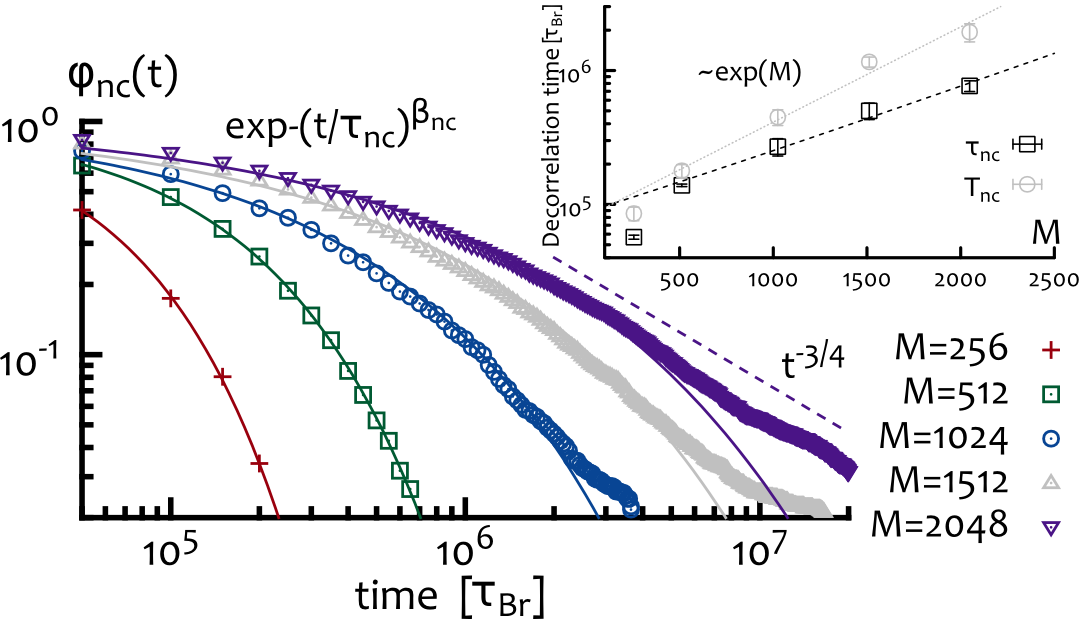}
\caption{ {\bf The rate at which chains become non-contiguous slows with their length}. Correlation function $\varphi_{\rm nc}(t)$ as a function of time (in simulation time units $\tau_{Br}$, see Materials and Methods) for different chain length. Solid lines represent stretched exponentials while the dashed line on the right represents a power law decay with exponent of $-3/4$ suggesting an even slower inter-ring de-correlation dynamics. The inset shows the exponential increase of the typical exchange time $\tau_{\rm nc}$ and $T_{\rm nc}$ for the systems displaying a stretched exponential relaxation (SI Appendix Fig.~S5).}
\label{fig:Main_PersNeigh}
\end{figure}

These findings are a strong signature that rings display long lasting inter-coil correlations that are present even after they have travelled beyond their own size, in agreement with previous numerical and experimental findings~\cite{Halverson2011a,Doi2015}. In addition, this is clear evidence that the exchange dynamics of the rings becomes slower, more glass-like~\cite{Lee2015,Kang2015}, as the polymerisation $M$ increases. The increasingly stretched decay of $\varphi_{\rm nc}$ also implies that the relaxation dynamics of the chains becomes more heterogeneous, \emph{i.e.}, some parts of the chains are much slower to separate from one another than other parts.
One might conjecture that the increasingly slow and heterogeneous exchange dynamics will eventually appear in the long-time dynamics of the ring displacement since contiguity will ultimately constrain motion.

It is natural to now ask whether we can understand the nature of these correlations, and the heterogenity in the relaxation dynamics observed in Fig.~2, as being directly related to the threadings recently proved to exist in concentrated solution of rings embedded in gels~\cite{Michieletto2014,Michieletto2014a}. In the case of pure solutions of rings, such threadings have never been rigorously identified.   

Our approach consists in artificially immobilising (``freezing''), in space and time, a fraction $c$ of randomly selected rings from equilibrated configurations and then tracking the dynamics of the ``un-frozen'' fraction. This protocol is inspired by recent theoretical and experimental studies on the nature of glass transitions and idealises the case in which a fraction $c$ of polymers in the system might be arbitrarily frozen, perhaps by mixing two polymeric species with different $T_g$ or by using optical tweezers~\cite{Gokhale2014,Nagamanasa2014}, although the primarily interest in this work is in its role as a conceptual tool.
 
If the rings are mutually threaded, perhaps in a way that resembles the threaded lattice animal in Fig.~\ref{fig:Panel_Snap_Threading}(c), then one would expect those un-frozen rings that are threaded by frozen ones to have their mobility substantially limited. They would appear to be immobilised within effective ``cages'', being the region of space that they can explore limited by the threadings that they experience. In the alternative picture where rings remain un-threaded, as is often envisaged in simplified models, the mobility would be substantially unaffected by the presence of frozen chains.

A primary result  of the present work, discussed below, is that we do indeed observe immobilisation of the un-frozen fraction. We believe that this represents excellent evidence for the existence of threadings in concentrated solutions of rings, something which has not previously been demonstrated.

\subsection{Randomly Pinning Rings Induces a Kinetically Arrested State}

Starting from an equilibrated configuration, we perturb the system by randomly freezing, in space and time, a fraction $c$ of coils. As a comparison, we first consider a system of linear polymers with one unfrozen linear ``probe'' chain diffusing through $cN=N-1$ artificially immobilised (frozen) linear polymers. The same perturbation is then applied to a dense solution of ring polymers, and the two cases are compared in Fig~3. This figure shows that the long-time dynamics of the unfrozen linear chain (green circles) is substantially insensitive to the presence of frozen neighbours. This is because the linear polymer can undergo reptation and simply snake through the frozen surroundings\footnote{There is a weak correlation effect due to the lack of mass relaxation in the frozen chains, leaving a ``hole'' (and corresponding ``bump") in the density as the mobile chain moves. We find that this only weakly affects the reptative dynamics.}.  On the other hand, when we repeated this procedure on a corresponding system of rings we observed the probe ring's diffusion to be arrested with it becoming irreversibly trapped within a region of space of size somewhat smaller than its gyration radius $R_g$ (see red squares in Fig.~3).

\begin{figure}[t]
 \centering
\includegraphics[width=0.48\textwidth]{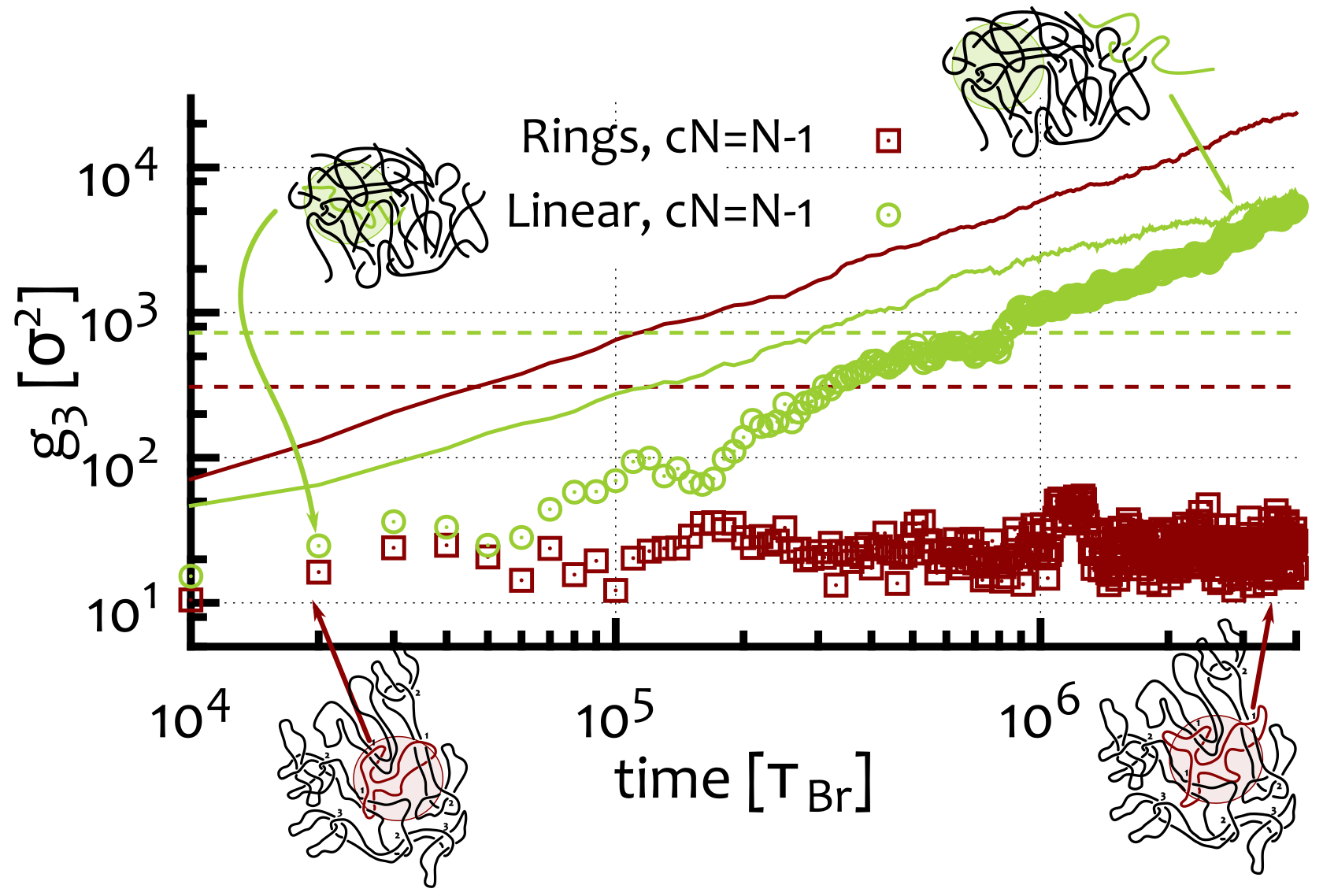}
\caption{{\bf A single un-frozen ring in a system of frozen chains becomes permanently caged while a linear polymer is substantially unaffected.}  Mean square displacement of the centre of mass $g_3(t)$ of a probe ring (red squares) and of a probe linear polymer (green circles) diffusing through, {\small $N-1$} frozen ring and linear polymers, respectively. The curves are averaged over 10 different probes (samples). The two systems have the same monomer density {\small $\rho=0.1\sigma^{-3}$} and are comprised of {\small $N=50$} polymers each containing {\small $M=256$} beads that differ only in their topology. Solid lines represent the behaviour of the free solution, \emph{i.e.} the {\small $c=0$} case, while dashed lines represent the diameter (squared) for rings and linear polymers, respectively (see Suppl. Movie M2).}
\label{fig:Freezing_MSD_Lin_vs_Rings}
\end{figure}

Because nothing other than the topology of the polymers was changed, this dramatically different dynamical response should be attributed to the presence of topological interactions between ring polymers, which we identify as the threadings. This immediately implies that the equilibrated state of the rings in our systems is one in which threadings constrain the free diffusion of the rings (see sketch in Fig.~1(c)), therefore limiting their motion and, in the extreme case, leading to caged diffusion when neighbouring rings that thread them are permanently frozen. The constraints provided by threading between rings in the unperturbed ($c=0$) solution will be transient to a greater or lesser extent but they must nonetheless exist. Our results therefore represent the first unambiguous evidence for inter-ring threading in dense solutions of rings.

\begin{figure*}[t]
\centering
\includegraphics[width=1.0\textwidth]{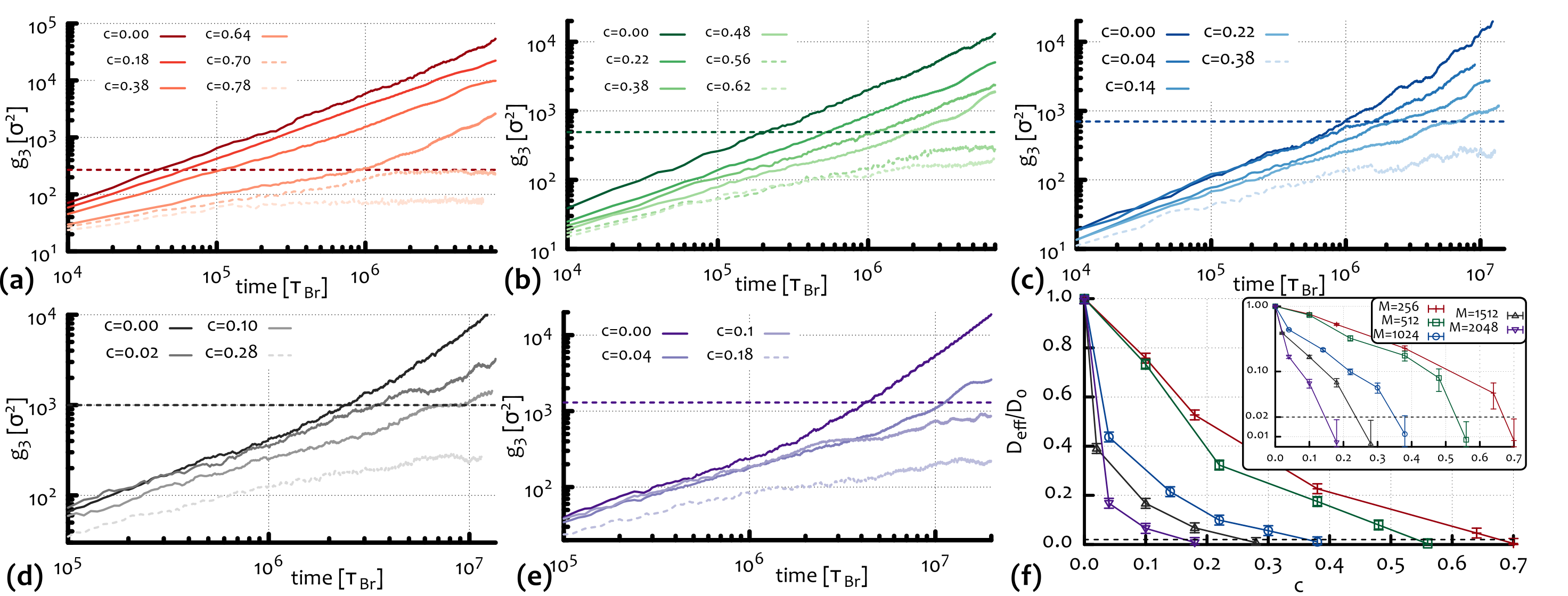}
\caption{{\bf The critical fraction of frozen chains $c^\dag$ at which all the non-explicitly frozen chains are usually caged by threadings decreases with ring length.} The mean squared displacement of the centre of mass ($g_3(t)$) averaged over only the non-frozen fraction ($1-c$) of rings plotted against time for different values of the frozen fraction $c$ ranging from $c=0$ (free solution) to $c \geq c^\dag$ at which all rings in the un-frozen fraction have become permanently caged. \textbf{(a)} $M=256$, \textbf{(b)} $M=512$ \textbf{(c)}$M=1024$, \textbf{(d)} $M=1512$ and \textbf{(e)} $M=2048$. 
The thick horizontal dashed lines indicate the diameter of the chains. \textbf{(f)} The value of $D_{\rm eff}$ (eq.~\eqref{eq:Deff}) normalized by $D_0\equiv D_{\rm eff}(c=0)$ is shown \red{in (main) linear-linear and (inset) linear-log scale to appreciate the nearly exponential dependence}. The critical value $c^\dagger$ is defined as the one for which $D_{\rm eff}(c^\dagger)/D_0$ is below a critical threshold that we here set to $0.02$ (dashed line). }
\label{fig:g3_liquidTOglass}
\end{figure*}
 
It is also worth stressing that the measurement that we perform should be interpreted as carrying a statistical meaning: With enough attempts one will always be able to find a ring that is not caged, however unlikely that might be. When we interrogate the rings regarding their state, caged or free, we are implicitly drawing from a binomial distribution and therefore we can calculate the probability $p$ of observing a non-caged ring in any one test for a given fraction of frozen rings. Having performed 10 tries freezing $N-1$ rings and having observed 10 caged rings, we can set a 95\% confidence bound on the fact that $p \simeq 0.26$, although the precise values are not particularly important and any statistical confidence criterion will give qualitatively similar results in what follows. 
 
Given that for all systems studied here we observe a regime in which all the un-frozen rings are caged in regions smaller than their sizes ($2R_g$) for the whole simulation run-time, it is natural to ask how the behaviour crosses over from the un-perturbed system ($c=0$), in which all the rings are free to diffuse and none are explicitly frozen, to the case in which enough rings are {\em explicitly} frozen to {\em implicitly} pin, or cage, the others (at some level of statistical confidence). 

We study this transition by tracking the behaviour of each chain's centre of mass diffusion $g_3(t)$, averaged over the unfrozen rings. The observed behaviour of $g_3(t)$ is reported in Fig.~\ref{fig:g3_liquidTOglass}: For all systems there exists a critical frozen fraction ($c^\dag$), for which every single unfrozen polymer is permanently trapped by the network of threadings.  In other words, at $c=c^\dag$, the systems exhibit a transition from (at least partially) liquid, or diffusive, behaviour to a glassy state in which the unfrozen chains, although free to re-arrange their conformations to some extent, are all irreversibly caged. 

More practically, one can define $c^\dag$ by introducing the effective diffusion coefficient~\cite{MCMarchetti2015}:
\begin{equation}
D_{\rm eff}(c) \equiv \lim_{t\rightarrow \infty} \dfrac{\langle g_3(t) \rangle}{6t}
\label{eq:Deff}
\end{equation}
whose average is taken over the unfrozen rings and vanishes when all the unfrozen rings are ``caged'' by the topological interactions. It is therefore natural to identify $c^\dag$ as the value of $c$ at which $D_{\rm eff} = 0$. In practice, we run the simulations for a time much longer than the relaxation time of the rings measured at $c=0$ (see also SI Appendix Fig.~S4) and define as ``caged'' all the systems that display a $D_{\rm eff}$ fifty times smaller than the diffusion coefficient of the unperturbed ($c=0$) case, $D_0$ (see Fig.~4(f)). From this figure one can notice that the decay of the $D_{\rm eff}$ becomes increasingly steeper as the rings become longer, suggesting that the systems with long chains are more susceptible to small contaminations of frozen chains \red{while from the inset one can also appreciate the nearly exponential decay of $D_{\rm eff}$ with $c$~\cite{Kob2014}.} 

While the diffusion coefficient of the centre of mass of the rings informs us about the overall diffusion of the chains, it is also interesting to study the relaxation of the chains at different length-scales in response to this external perturbation. This can be done by computing the dynamic scattering function
\begin{equation}
S_c(q,t) = \dfrac{1}{S_c(q)}\left\langle \dfrac{1}{fN} \sideset{}{'}\sum_{I} \left(\dfrac{1}{M} \sum_{ij \in I} e^{i \bm{q}\left[ \bm{r}_i(t) - \bm{r}_j(0)\right]} \right) \right\rangle, 
 \label{eq:scatteringfunction}
\end{equation}
where $\sum^\prime$ stands for the summation over the $(1-c)N=fN$ non-frozen chains and the average is performed over time and orientations of $\bm{q}$.
In Fig.~5 (and SI Appendix Figs.~S6-S8) we report the behaviour of this quantity computed for two choices of wave vector $q$ probing length scales ($l=2\pi/q$) comparable to the rings' diameter $2R_g$. We observe that $S_c(2/R_g,t)$ decays much more slowly than the scattering function measured at $q=4/R_g$. In order to compare their behaviour we choose an arbitrarily long time ($\bar{t}=10^7$ $\tau_{Br}$) at which we evaluate the scattering function and report their difference $\Delta S_c$ for a range of values of $c$ and rings' length (see Fig.~5, and see SI Appendix for different choices of $\bar{t}$).

\begin{figure}[h]
\vspace*{-0.4 cm}
\centering
\includegraphics[width=0.5\textwidth]{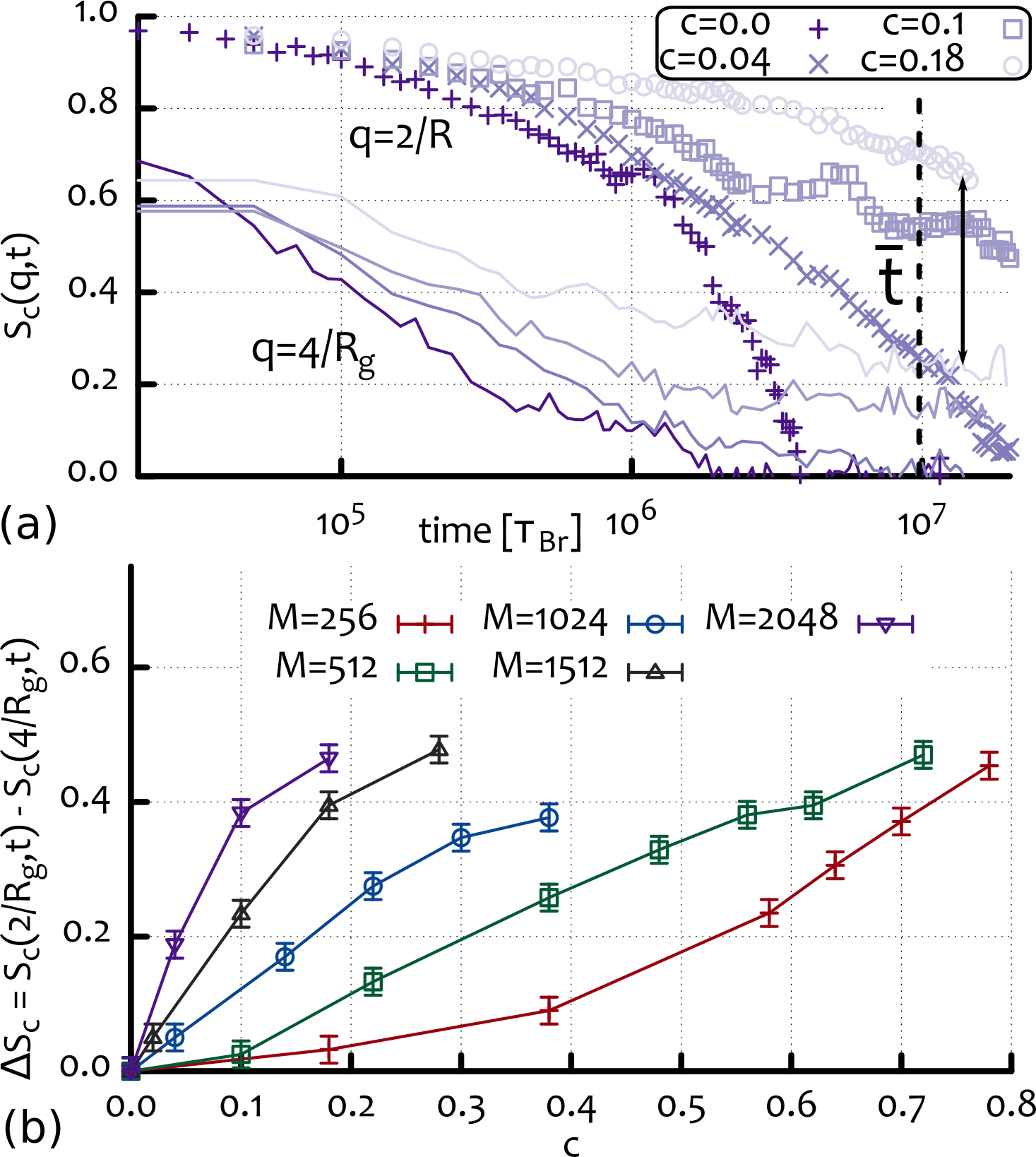}
\caption{{\bf The freezing procedure affects long length scales more strongly than short ones.} \textbf{(a)} Behaviour of the dynamic scattering function $S_c(q,t)$ for the system with $M=2048$ and for $q=2/R_g$ ($l=\pi R_g$) (symbols) and $q=4/R_g$ ($l=\pi R_g/2$) (solid lines) (more examples can be found in SI Appendix Figs.~S7 and S8). The difference $\Delta S_c=S_c(2/R_g,\bar{t})-S_c(4/R_g,\bar{t})$ is computed at an arbitrary long time $\bar{t}=10^7$ $\tau_{Br}$ and reported in \textbf{(b)} for the different cases. From this plot it is clear that the relaxation of length scales longer than $2R_g$ is slowed down more severely than the relaxation of shorter ones by the pinning procedure. In turn, this suggests that threadings act mainly by constraining low wavenumber modes.}
\label{fig:NonErg}
\vspace*{-0.4 cm}
\end{figure}

The increasing trend of $\Delta S_c$ suggests that length scales longer than the diameter of the rings are more susceptible to the freezing procedure. This is consistent with the fact that threadings between ring polymers constrain their translational degrees of freedom on length scales comparable to the size of the rings while there can exist internal modes that are left un-hindered and free to relax. In addition, we observe that the systems with longer chains require a smaller value of $c$ to significantly slow down the relaxation of the scattering function with respect to systems with shorter chains. This observation is in turn in agreement with the fact that threadings are more numerous for systems with longer chains~\cite{Michieletto2014}, and therefore a smaller fraction of frozen rings is sufficient to achieve a similar slowing down.

Fig.~5 also suggests that while coils are strongly constrained by the presence of frozen rings, the configurations can still partially relax by internal re-arrangements and this allows rings to relax their internal stress. The picture that emerges is rather different than that for linear polymers, where the centre of mass diffusion is intimately related to its ability to undergo any and all conformational re-arrangements. In the latter case, motion is only arrested by the onset of a microscopically glassy state for $T<T_g$. In the case of ring polymers, given their closed topology, threadings instead decouple the dynamics involved in displacing the centre of mass and those involved in the internal re-arrangement. Any glassy jamming that emerges in this work is unusual in that it is not associated with arrest of microscopic degrees of freedom but rather the topological arrest of low wavenumber modes, including that associated with the motion of the centre of mass. This is the reason why an entanglement plateau is absent. It also means that the slowing down of the dynamics in systems of rings might not be clearly captured by the stress relaxation function $G(t)$ frequently studied in the literature~\cite{Kapnistos2008,Pasquino2013,Doi2015,Bras2014}, which is mostly dominated by the unconstrained internal modes, but rather by the mean square displacement of the unfrozen rings (as in Fig.~\ref{fig:g3_liquidTOglass}) or the dynamic scattering function $S_c(q,t)$.

\vspace*{-0.4 cm}
\subsection{The Phase Diagram of the System}

The behaviour of both, the effective diffusion coefficient $D_{\rm eff}$ and $\Delta S_c$, show an increasingly steeper dependence on the freezing parameter $c$ as the length of the rings increases (see Fig.~4(f) and 5(b)). This implies that systems made of long chains become extremely sensitive to very small perturbations. Indeed one can think of these observables as quantifying a form of susceptibility that captures how the dynamic mobility of the system responds to the freezing of (very few) threading constraints by chain immobilisation. 

\begin{figure}[h]
   \vspace*{-0.2 cm} 
	\centering
	\includegraphics[width=0.5\textwidth]{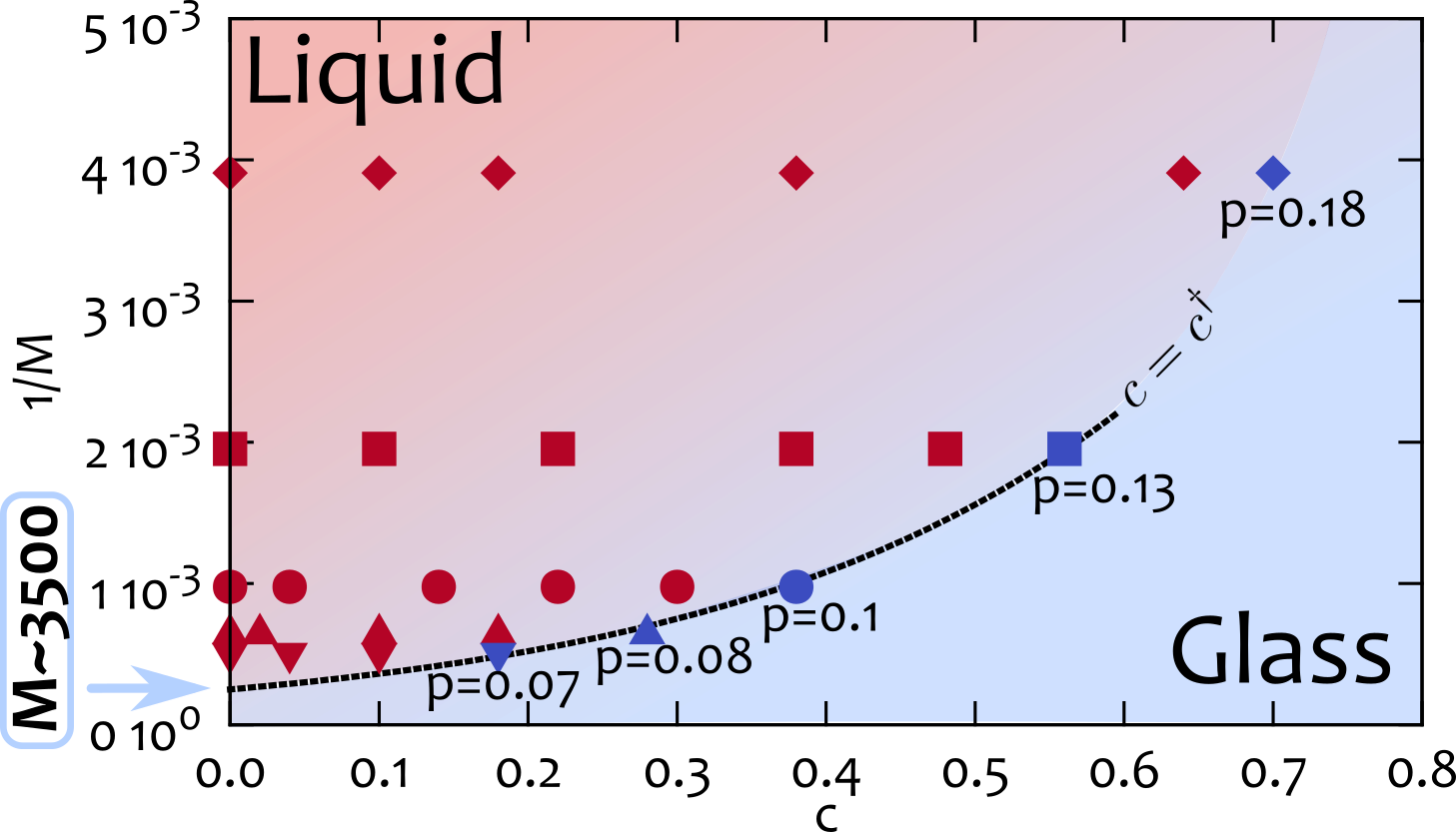}
	\caption{{\bf The phase diagram of the system suggests a spontaneous vitrification at large $M$.}
		Phase space {\small $(1/M,c)$} for systems of rings with length {\small$M$} and a fraction {\small$c$} of rings permanently frozen in space and time. The transition line {\small$(1/M^\dag,c^\dag)$} is shown together with an exponential fit (dotted line). The coloured data points in the diagram indicate whether the system displays a finite diffusion coefficient at large times (red) or whether it is irreversibly caged with vanishing $D_{\rm eff}$ (blue). Along the transition line we also report the value of the probability $p$ of finding an un-caged ring in any one test performed at fixed $c$ (see text for details). }
	\label{fig:PhaseDiagram}
	\vspace*{-0.4 cm} 
\end{figure}

Finally, in Fig.~6 we show the transition line $c=c^\dag$ in the space of parameters $(1/M,c)$. The coloured data points represent the position of the simulated systems in the phase space and whether their behaviour was liquid-like (finite $D_{\rm eff}$ --  red) or solid-like (vanishing $D_{\rm eff}$ -- blue). 
The diagram in Fig.~\ref{fig:PhaseDiagram} is reminiscent of that observed in more traditional glass-forming systems subject to random pinning fields~\cite{Cammarota2012} where the temperature $T$ is here replaced by the inverse length $1/M$; a substitution not unfamiliar to field-theoretic treatments of polymer systems~\cite{Gennes1979}.
Along the transition line we also report the value of the probability $p$ of finding an un-caged ring in any one sample as obtained from the binomial distribution at 95\% confidence interval $(1-p)^{(1-c^\dagger)N}=0.05$.
It is also worth noticing that the curve is well fitted by an exponential function of the form
$ M^\dag(c) = M_g e^{- 3.3 c}$ with $M_g \simeq 3500$. Computational limitations forbid a thorough exploration of the small $c$ region; Nonetheless, and somewhat remarkably, our results show that the number of caged chains per explicitly frozen ring is exponential in the ring length (see SI Appendix Fig.~S9(b)). This would seem to mean that an arbitrarily large fraction of caged chains can be achieved from an arbitrarily small fraction of frozen rings provided that the rings are long enough. This raises the possibility that a glassy state could emerge spontaneously in the critical regime near $c=0$ in the (universal) limit of large $M$.

From these results it is clear that concentrated solutions of long rings are extremely sensitive to a small external pinning field that can drive the centre of mass motion of the rings to become glassy (jammed). We have here traced this phenomenon to the unusual, topological constraints provided by the threadings between ring polymers. While branched polymers also display exponentially-long relaxation times, the origin of this is not {\em cooperative} in nature and would therefore be rather insensitive to the freezing of some components. On the other hand, cooperativity, such as the inter-ring threadings we study, is thought to be an essential ingredient of a genuine glass transition~\cite{Gibbs1965}. 

Our work therefore elucidates the conformation of rings in concentrated solutions and unambiguously characterises, for the first time, the presence of threadings between rings. Furthermore, we show the dynamics of ring polymers to be sensitive to these interactions and, finally, we provide strong evidence for the emergence of a kinetically arrested state solely driven by topological constraints. 
%since both density and temperature are kept constant in the systems. 
This system is therefore a novel instance of a glassy state induced through the topology of the constituents.

\begin{materials}
\subsection{Simulations Details}
We model the system by enclosing {\small $N$} semi-flexible polymers modelled using  the standard Kremer-Grest model~\cite{Kremer1990} and formed by {\small $M$} beads of nominal size {\small $\sigma$} in a box with periodic boundary conditions of linear size {\small $L$}. The monomer fraction is fixed at {\small $\rho=NM/L^3=0.1\sigma^{-3}$} and we vary the length of the polymers {\small $M$} as our main control parameter. The simulation time-step is denoted with {\small $\tau_{Br}$} and correspond to the time taken by a bead of size {\small $\sigma$} experiencing a friction {\small $\xi$} to diffuse its size. The simulation is carried out in NVT ensemble with the LAMMPS engine. Further computational details are described in the SI Appendix. 
\end{materials}

\begin{acknowledgments}
The authors would like to thank Mike Cates, Davide Marenduzzo, Enzo Orlandini, Christos Likos and Dimitris Vlassopoulos for useful discussions and careful remarks throughout the project. We also acknowledge the support of EPSRC through the Complexity Science Doctoral Training Centre at the University of Warwick (DM), grant EP/E501311 and a Leadership Fellowship (MST), grant EP/I005439/1. 
The computing facilities were provided by the Centre for Scientific Computing of the University of Warwick with support from the Science Research Investment Fund. 
The authors also gratefully acknowledge the computing time granted by the John von Neumann Institute for Computing (NIC) and provided on the supercomputer JURECA at Jülich Supercomputing Centre (JSC).
\end{acknowledgments}

%\vspace*{- 0.5 cm}

\appendix[Supplementary Information]
\setcounter{figure}{0}
\makeatletter 
\renewcommand{\thefigure}{S\@arabic\c@figure}
\makeatother

\subsection{Computational Details}
We model the system by enclosing \red{$N=50$} semi-flexible bead-spring polymers formed by $M$ beads in a box with periodic boundary conditions of linear size $L$. The monomer density $\rho=NM/L^3=0.1\sigma^{-3}$ is fixed and we vary the length of the polymers and the size of the box \red{to keep the density constant}. The chains are modelled via the Kremer-Grest worm-like chain model~\cite{Kremer1990}.
as follows:
Let $\bm{r}_i$ and $\bm{d_{i,j}} \equiv \bm{r}_j - \bm{r}_{i}$  be respectively the position of the center of the $i$-th bead and the vector of length $d_{i,j}$ between beads $i$ and $j$, the connectivity of the chain is treated within the finitely extensible non-linear elastic model with potential energy, 
\begin{equation}
U_{FENE}(i,i+1) = -\dfrac{k}{2} R_0^2 \ln \left[ 1 - \left( \dfrac{d_{i,i+1}}{R_0}\right)^2\right]  \notag
\end{equation}
for  $d_{i,i+1} < R_0$ and $U_{FENE}(i,i+1) = \infty$, otherwise; here we choose $R_0 = 1.6$ $\sigma$ and $k=30$ $\epsilon/\sigma^2$ and the thermal energy $k_BT$ is set to $\epsilon$. 
The bending rigidity of the chain is captured with a standard Kratky-Porod potential,
\begin{equation}
U_b(i,i+1,i+2) = \dfrac{k_BT l_p}{2\sigma}\left[ 1 - \dfrac{\bm{d}_{i,i+1} \cdot \bm{d}_{i+1,i+2}}{d_{i,i+1}d_{i+1,i+2}} \right],\notag
\end{equation}
where we set the persistence length $l_p = 5 \sigma$. The steric interaction between beads is taken into account by a truncated and shifted Lennard-Jones (WCA) potential  
\begin{equation}
U_{LJ}(i,j) = 4 \epsilon \left[ \left(\dfrac{\sigma}{d_{i,j}}\right)^{12} - \left(\dfrac{\sigma}{d_{i,j}}\right)^6 + 1/4\right] \theta(2^{1/6}\sigma - d_{i,j}) \notag.
\end{equation} 
where $\theta(x)$ is the Heaviside function.

Denoting by $U$ the total potential energy, the dynamic of the beads forming the rings is described by the following Langevin equation:
\begin{equation}
m \ddot{\bm{r}}_i = - \xi \dot{\bm{r}}_i - {\nabla U} + \bm{\eta}
\label{langevin}
\end{equation}
where $\xi$ is the friction coefficient and $\bm{\eta}$ is the stochastic delta-correlated noise. The variance of each Cartesian component of the noise, $\sigma_{\eta}^2$ satisfies the usual fluctuation dissipation relationship $\sigma_{\eta}^2 = 2 \xi k_B T$.

As customary~\cite{Kremer1990}
we set $m/\xi = \tau_{LJ}=\tau_{Br}$, with the LJ time $\tau_{LJ} = \sigma \sqrt{m/\epsilon}$ and the Brownian time $\tau_{Br}=\sigma/D_b$, where $D_b = k_BT/\xi$ is the diffusion coefficient of a bead of size $\sigma$, is chosen as simulation time step. From the Stokes friction coefficient of spherical beads of diameters $\sigma$ we have: $\xi = 3 \pi \eta_{sol} \sigma$ where $\eta_{sol}$ is the solution viscosity. 
It is possible to map this to real-time units by using the nominal water viscosity, $\eta_{sol}=1$ $cP$ and setting $T=300$ K and $\sigma$ equal, for instance,to the diameter of hydrated B-DNA ($\sigma=2.5$ $nm$), for which one has $\tau_{LJ} = \tau_{Br} = {3 \pi \eta_{sol} \sigma^3/\epsilon} \simeq 37$ $ns$. 
The numerical integration of Eq.~\eqref{langevin} is performed by using a standard velocity-Verlet algorithm with time step $\Delta t = 0.01 \tau_{Br}$ %\sim 0.4$ $ns$ 
and is implemented in the LAMMPS engine.

\red{
	\subsection{System Preparation and Equilibration}
	The systems are prepared by placing the rings randomly in a very large box. The linking number between all pairs of rings is also checked in order to avoid linked polymers. In addition, the rings are initialised as perfect circles in order to avoid self-knotting.
	The desired monomer density is achieved by slowly shrinking the box until the target box size is reached (effectively applying a constant pressure). At this stage, we checked for unwanted linked rings and found none. After this, we equilibrate the systems by performing standard runs (with no rings artificially pinned) for at least the time need for the chains to displace their centres of mass of several $R_g$'s. We observe that $t=10^7 \tau_{Br}$ time-steps are enough to obtain this condition. After the equilibration we performed another run in order to study the free, \emph{i.e.} unperturbed, behaviour of the system. The mean square displacement obtained from this run is reported in see Fig.~S4. The simulations in which we artificially pin some of the rings are then started from the late stages of this last run, so that the initial configuration for these \emph{perturbed} simulations were un-correlated from the initial system set up.
	% the un-correlation time $R^2_g/D_{CM}$ and the results obtained were independent of the initial state. 
	The rings that are artificially pinned are chosen at random among the $N$ rings. Because the simulations are very computationally expensive, we only perform one simulation for each choice of $c$. For the longest chains reported here ($M=2048$ beads) each run up to $2$ $10^7$ $\tau_{Br}$ takes up to 4 weeks when running in parallel over 64 processors. This time-window has to be run for every choice of the fraction of pinned chains $c$. As we tested four choices of $c$, the results reported only for the system with $M=2048$ (Fig.~3(e) in main text) take $\sim 4$ months of $64$ CPUs time or, equivalently, $\sim 20$ years of single CPU time.
}

\subsection{The Size and Static Structure of Rings are in Agreement with the Crumpled Globule Behaviour}
\begin{figure*}[t]
	\centering
	\includegraphics[width=0.9\textwidth]{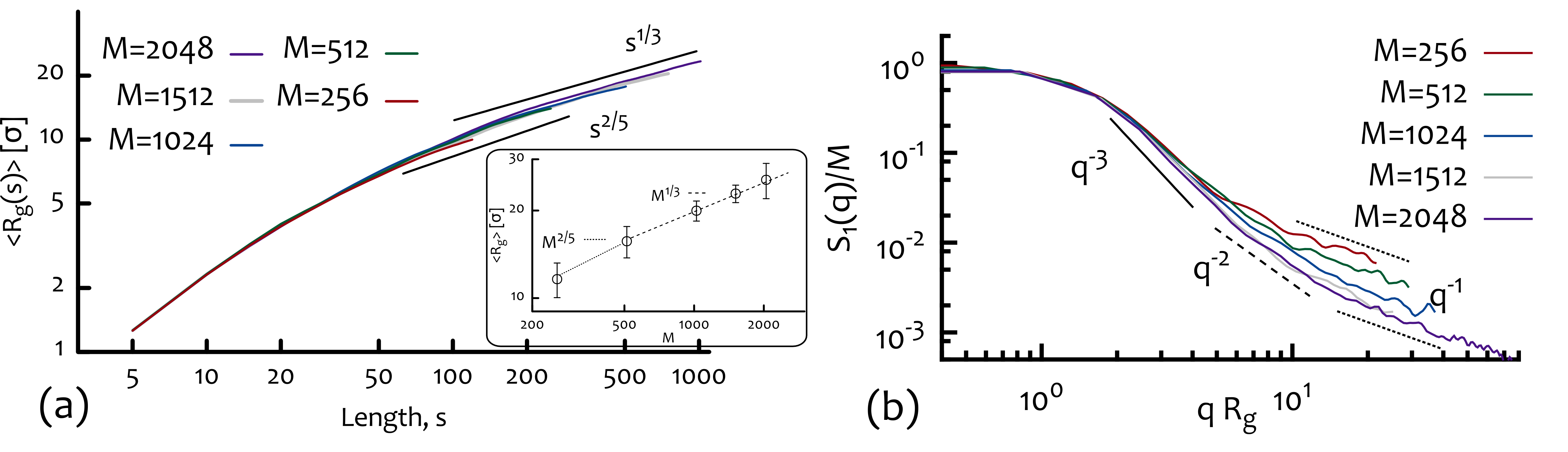}
	\caption{(a) \red{Ensemble average of the radius of gyration $\langle R_g \rangle \equiv \langle R^2_g \rangle^{1/2}$ and plotted against rings' contour length $s$ (main) and total length $M$ (inset) both in units of the bead size. As reported in the literature, the rings show a collapsed behaviour for large $M$}. (b) Static scattering function $S_1(q)$ plotted against $qR_g$ and normalised by the length of the rings $M$. This quantity indicates a complex arrangement of the rings internal structure, which does not seem to follow a unique fractal dimension at all lengths. }
	\label{fig:SI_Rg_SqSt}
\end{figure*}

In agreement with results reported in the literature~\cite{Halverson2011,Halverson2011a} we observe (Fig.~S1) that the radius of gyration of the rings scales as $R_g\sim M^{\nu}$ with \red{$\nu\simeq 1/3$} in the limit of large polymerisation index $M$, while we observe \red{$\nu \simeq 2/5$} for shorter rings. \red{This is supported by the measurement of $\langle R_g^2 \rangle$ either for the whole rings or as a function of the contour length $s$. The values of the exponents are in agreement with previously reported findings and we refer to previous works\cite{Halverson2011,Halverson2011a} for dedicated measurements of $\nu$.}
Another way of investigating the conformation of the rings is by measuring the static structure factor. 
For wave-vectors in the range $1/R_g < q < 1/\sigma$, one should expect that $S_1(q)$ defined as
\begin{equation}
S_1(q) =   \left\langle \dfrac{1}{M} \sum_{i,j \in I}^M e^{i \bm{q} (\bm{r}_i(t) - \bm{r}_j(t))} \right\rangle
\label{eq:StaticFactor}
\end{equation}
where the indexes $i$,$j$ run over ring $I$, to give $S_1(q) \sim q^{-D_F}$~\cite{deGennes1979}, where $D_F$ is the fractal dimension of the chain at length scale $1/q$ and it is related to the scaling exponent $\nu$ as $D_F = 1/\nu$.  Linear chains in the melt display $D_F=2$ for a broad range of $q$'s~\cite{Kremer1990}  
while we observe the rings to have a more complex organisation with $D_F$ ranging from $D_F\simeq 3$ to $D_F\simeq 1$ at large $q$, in agreement with previous findings~\cite{Halverson2011}
(see Fig.~\ref{fig:SI_Rg_SqSt}).

%Figure R_g and S_q static

\begin{figure*}[t]
	\centering
	\includegraphics[width=0.9\textwidth]{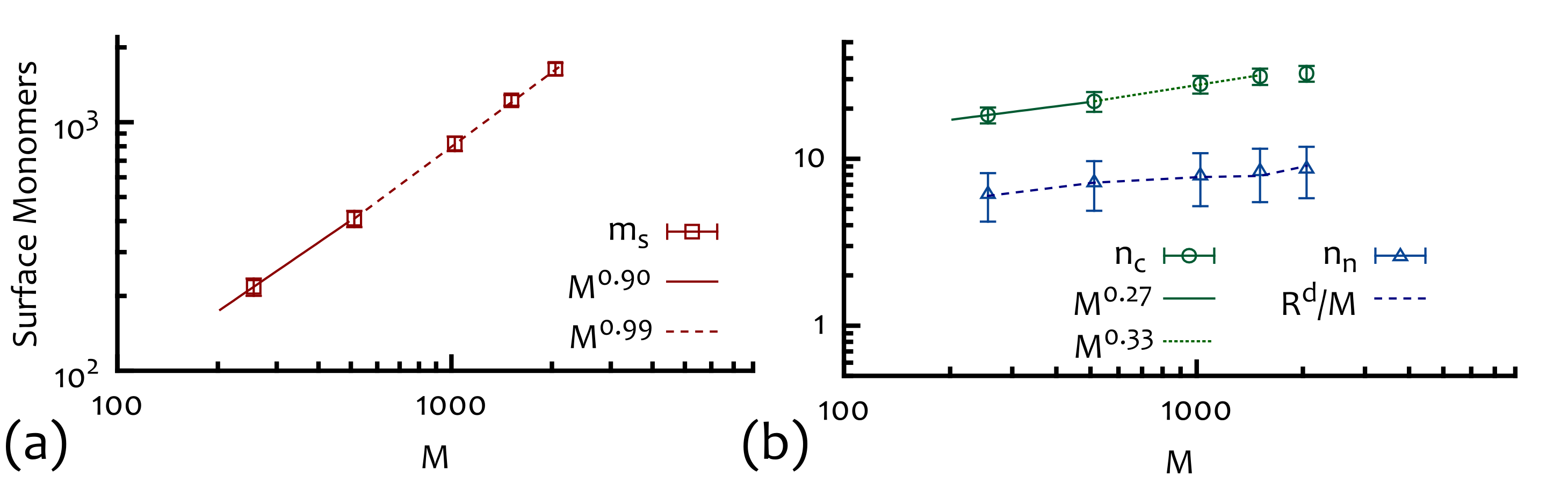}
	\caption{(a) The number of surface monomers $m_s$ shows a scaling $m_s \sim M^{\beta_c}$ with $\beta_c \simeq 0.99 $ for the longest chains. (b) The number of contiguous chains $n_c$ and number of neighbouring chains $n_n$ is shown. While their scaling behaviour is similar, the average number of chains in contact at any time is systematically larger than the number of chains whose centre of mass is closer than $2R_g$. This may imply large fluctuations in the rings conformations that bring distant chains in contact with one another. \red{In (b) we show two curves with exponents $0.33$ and $0.27$ as a guide for the eye but refer to the literature and further studies for more precise estimates.}}
	\label{fig:SI_SurfaceMonomers}
\end{figure*}

\subsection{The Contact Surface of the Coils Grows Extensively with the Length of the Rings}
In order to quantify the degree of interaction between coils, we investigate (i) the number of surface monomers, (ii) the number of contiguous chains and (iii) the number of neighbouring chains. 
\begin{itemize}
	\item[](i) The number of surface monomers $m_s$ is computed by counting the number of beads forming the chains that are in contact with beads forming any other chains, according to the contact matrix in eq.~3 of the main text, \emph{i.e.} any two beads are in contact if their position is closer than $d = \rho^{-1/3}$, where $1/\rho=10\sigma^3$ is the free volume per bead.
	
	\item[] (ii) The number of contiguous chains $n_c$ is computed the number of chains that have surface beads that are in contact. 
	
	\item[] (iii) The number of neighbouring chains $n_n$ is instead defined as the number of coils that are closer than $2Rg$ to any one other coil. 
\end{itemize}
These quantities are reported in Fig.~\ref{fig:SI_SurfaceMonomers}. \\
The surface monomers show a near extensive dependence to the length of the rings, as already observed in previous works~\cite{Halverson2011}, while the number of contiguous and neighbouring chains show a similar scaling behaviour as a function of $M$, although $n_c$ is found systematically larger than $n_n$. This may imply large fluctuations in the rings conformations, which bring distant coils in contact.

\subsection{The Contact Probability Shows a Decay Consistent with the Mean-Field Estimate $\gamma=1$}

The contact probability is defined as
\begin{equation}
P_c(|i-j|) = \left\langle\dfrac{1}{M}\sum_{i=1}^{M-1} \sum_{j=i+1}^M \Theta(a - |\bm{r}_i(t) - \bm{r}_j(t)|) \right\rangle
\label{eq:Pc}
\end{equation}
where $\Theta(x)$ is the Heaviside function and $a$ is the chosen cut-off. In Fig.~S3 we report $P_c$ for two value of $a$ and for different chain lengths. The behaviour of $P_c(m)$ is expected to follow the crumpled globule scaling 
\begin{equation}
P_c(m) \sim m^{-\gamma}
\end{equation}
with $\gamma \gtrsim 1$, and for which the mean field value $\gamma=\nu d$ is a lower bound. We here observe $\gamma\simeq 1.02 - 1.09$ (see Fig.~\ref{fig:SI_Gr}). The prediction that fixes the sum of the contact exponent $\gamma$ and the surface exponent $\beta_c$ equal to 2 in the case of crumpled globules ($\nu=1/3$), \emph{i.e.}  $\beta_c+\gamma=2$~\cite{Halverson2014}, is therefore here recovered within errors.

\subsection{The Pair Correlation Function Suggest that the Coils are Largely Inter-Penetrating}
In order to probe the inter-penetration of the coils one can also investigate the pair correlation function $g(r)$ which we here defined similarly to a recent work~\cite{Kang2015}
\begin{equation}
g(r) = \dfrac{2}{N(N-1)}\sum_{I=1}^{N-1} \sum_{J=N+1}^N \delta[|\bm{r}_{CM,I}(t) - \bm{r}_{CM,J}(t)| - r]
\label{eq:si_gr}
\end{equation}
where $\bm{r}_{CM,I}$ indicates the position of the centre of mass of ring $I$. \\

\begin{figure*}[t]
	\centering
	\includegraphics[width=0.95\textwidth]{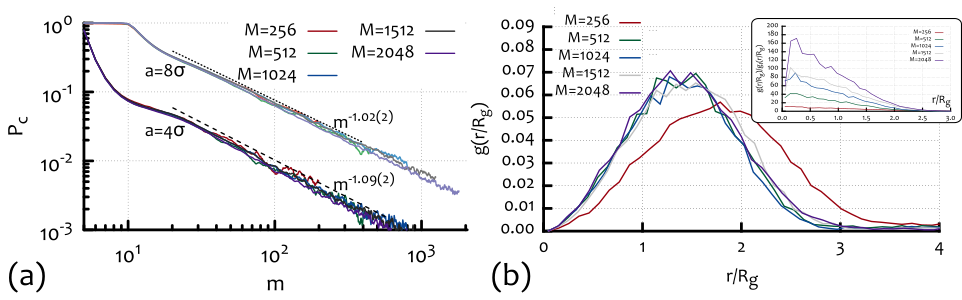}
	\caption{(a) The contact probability function $P_c(m)$ (defined in eq.~\eqref{eq:Pc}) shows a scaling behaviour $m^{-\gamma}$ with $\gamma$ slightly dependent on the choice of the cut-off $a$ but compatible with previous findings~\cite{Halverson2011}. (b) Pair correlation function $g(r)$ for the rings centre of mass as defined in eq.~\eqref{eq:si_gr}. The peak position imply that the coils are largely overlapping. \red{In the inset we report $g(r)$ normalised by the ideal pair distribution function $g_I(r)=4\pi\rho_c r^2$, where $\rho_c \equiv N/L^3$ is the coils' density. The coils behave as ultra-soft sphere with large inter-penetrations.} }
	\label{fig:SI_Gr}
\end{figure*}

\red{This function has been used in a recent work~\cite{Kang2015} probing the glassy dynamics of polymers under confinement and we here find well characterising the degree of overlap between coils. The behaviour of $g(r)$ (reported in Fig.~\ref{fig:SI_Gr}) in fact shows a distinct peak at $r_{\rm c}\simeq 1.8R_g$ for $M=256$ and at $r_{\rm c}\simeq 1.4R_g$ for $M\geq 512$. This implies that the coils, although crumpled, are strongly overlapping.} \\

\begin{figure*}[t]
	\centering
	\includegraphics[width=0.8\textwidth]{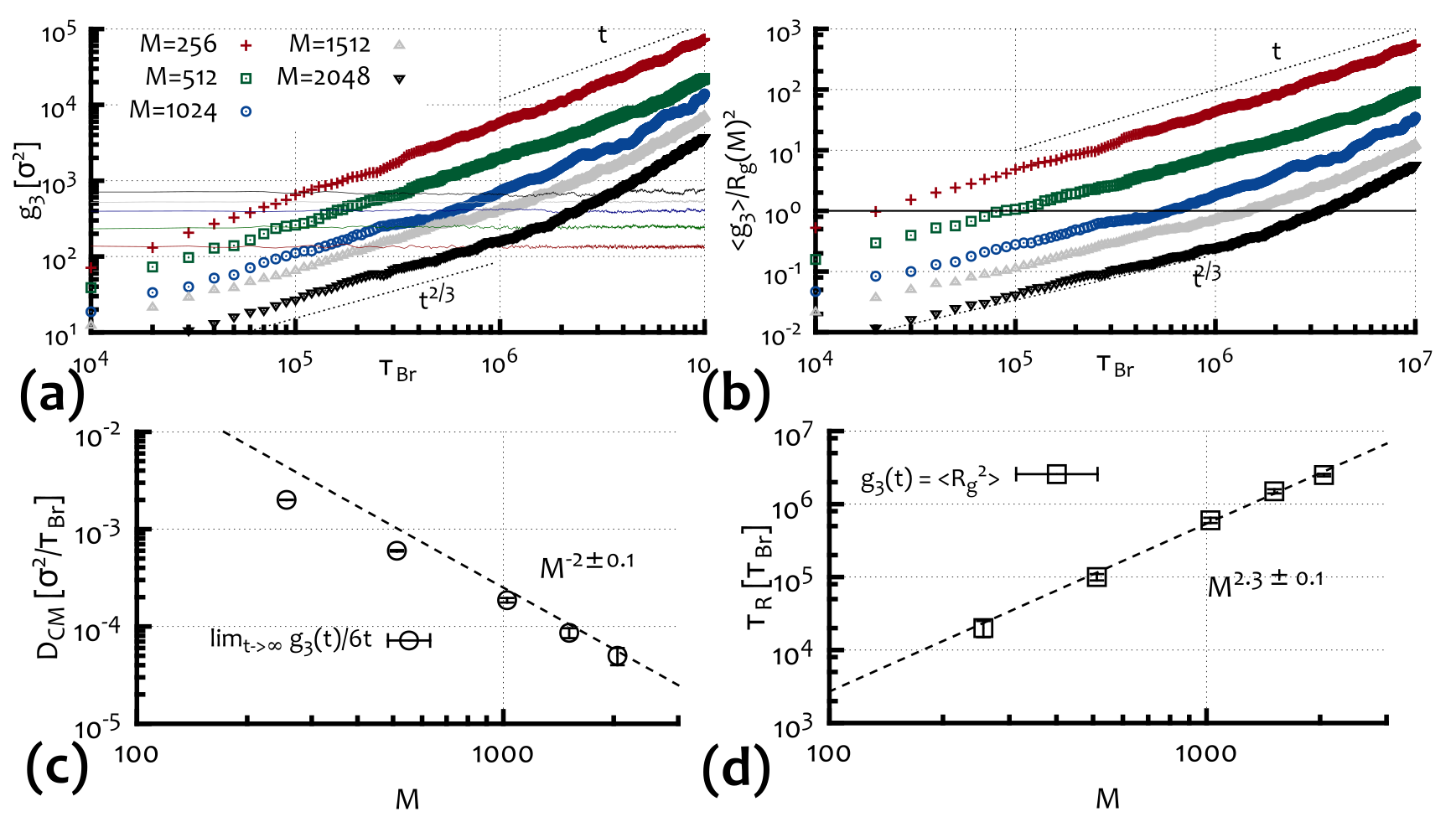}
	\caption{\textbf{(a)} $g_3(t)$: Mean square displacement (MSD) of the rings centre of mass. The faint horizontal lines represent the square radius of gyration $\langle R^2_g (t)\rangle$. \textbf{(b)} $g_3(t)/R_g^2$: MSD of the rings centre of mass divided by their squared gyration radius. The solid horizontal line in \textbf{(b)} marks the value $g_3(t)/R_g^2=1$ at which the rings have, on average, travelled once their own size. \textbf{(c)} and \textbf{(d)} Report the behaviour of the diffusion coefficient of the centre of mass of the rings and the relaxation time $\tau_R$ (see eq.~\eqref{eq:tau_relax}), respectively. \red{By fitting the last three data-points in order to obtain the asymptotic values of $D_{CM}$ for large $M$ we obtain $-2 \pm 0.1$. By considering all data point we obtain a value for the exponent of $D_{CM}$ of around $1.8 \pm 0.1$.} }
	\label{fig:SI_MSD}
\end{figure*}

\newpage
\subsection{The Mean Squared Displacement of the Unperturbed System is in Agreement with Previous Observations}
In Fig.~\ref{fig:SI_MSD} we report the rings centre of mass mean square displacement $g_3(t)$ defined as 
\begin{equation}
g_3(t) = \left\langle \left[ \bm{r}_{CM}(t) - \bm{r}_{CM}(0) \right]^2 \right\rangle,
\end{equation}
along with the diffusion coefficient $D_{CM} \equiv \lim_{t\rightarrow \infty} g_3(t)/6t$ and the relaxation time $\tau_R$ defined via the following condition
\begin{equation}
g_3(\tau_R) \equiv R^2_g.
\label{eq:tau_relax}
\end{equation}
As one can notice, the mean square displacement (MSD) of the centre of mass displays an intermediate sub-diffusive regime in which $g_3(t) \sim t^{3/4}$ before crossing over to a diffusive regime at large times. This is most evident for longer rings. 
The scaling of the diffusion coefficient as a function of the rings length is comparable to the one found in Ref.~\cite{Halverson2011a} although slightly smaller, which is in agreement with the lower monomer density considered in this work. This scales \red{asymptotically} as 
\begin{equation}
D_{CM} \sim M^{-2}
\end{equation} 
as well as the relaxation time $\tau_R$ for which we find
\begin{equation}
\tau_R \sim M^{2.3}.
\end{equation}
%This is most likely due to the fact that our system is sparser than the one studied in the literature, in which the monomer fraction is set to  $\rho=0.85$.
%which is comparable with the crude estimate (using $\nu=1/3$) $\tau_R \sim R^2_g/D_{CM} \simeq M^{2.7}$ 

\subsection{Persistent Contiguous Chains show an Exponentially Slow Uncorrelation Time}
\begin{figure*}[t]
	\centering
	\includegraphics[width=0.8\textwidth]{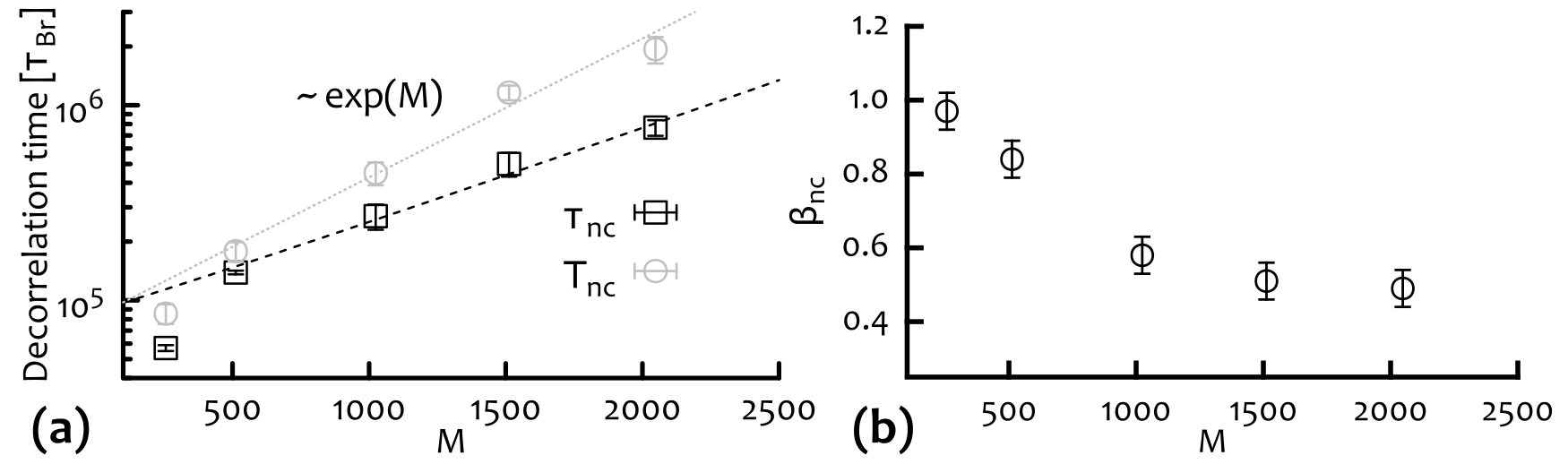}
	\caption{%\textbf{(a)} Persistent contiguous chains $\varphi_{nc}$ as defined in eq.~\eqref{eq:PersNeigh}; solid lines represent stretched exponentials $\varphi \sim \exp{-\left(t/\tau_{nc}\right)^\beta_{nc}}$, while the dashed line on the right is guide for the eye and indicates a power law of $t^{-0.75}$. $\beta\simeq 0.5$ and the dashed line a power law $t^{-3/2}$ 
		\textbf{(a)} Values of the relaxation times $\tau_{nc}$ \red{and $T_{\rm nc}$} for the contiguous correlation function $\varphi_{nc}(t)$ reported in Fig.~2 of the main text.%\ref{fig:Main_PersNeigh}. 
		We observe an exponential increase of the typical exchange time for large $M$. \textbf{(b)} Value of the stretching exponent $\beta_{nc}$, ranging from near unity for $M=256$ to around $1/2$ for $M=2048$.}%\ref{fig:Main_PersNeigh}.}
	\label{fig:SI_PersContig}
\end{figure*}

In the main text we report the behaviour of the correlation function $\varphi_{\rm nc}(t)$, characterising the exchange dynamics of the coils. 
In Fig.~\ref{fig:SI_PersContig} we report the values of the relaxation time of the exchange dynamics $\tau_{\rm nc}$ and the value of the stretching exponent $\beta_{nc}$ used to fit the data to stretched exponentials of the form 
\begin{equation}
\varphi_{\rm nc} = \exp{\left[-\left(\dfrac{t}{\tau_{nc}}\right)^{\beta_{nc}}\right]}.
\label{eq:SI_Stretchexp}
\end{equation}
\red{We also report the value of $T_{\rm nc}$, which is here defined as 
	\begin{equation}
	T_{\rm nc} \equiv \int_0^\infty \varphi_{\rm nc}(t) dt.
	\end{equation}	
	Both relaxation times $\tau_{nc}$ and $T_{\rm nc}$ are} observed to grow exponentially in $M$. The stretching parameter $\beta_{nc}$ is found to reach values close to $1/2$ for the longest chains studied in this work. This implies that the exchange time of the rings becomes extremely slow in the limit of large $M$ and in turn this may suggest the onset of a glassy dynamics (see discussion of Fig.~2 in the main text).\\

%It is also worth noting that for the system with longest chain ($M=2048$) we observe a significant deviation from the stretched exponential behaviour that can be fitted by a power law with exponent close to $-0.75$, suggesting an even stronger persistence of the inter-coil correlations at large times.
\begin{figure*}[t]
	\centering
	\includegraphics[width=1.0\textwidth]{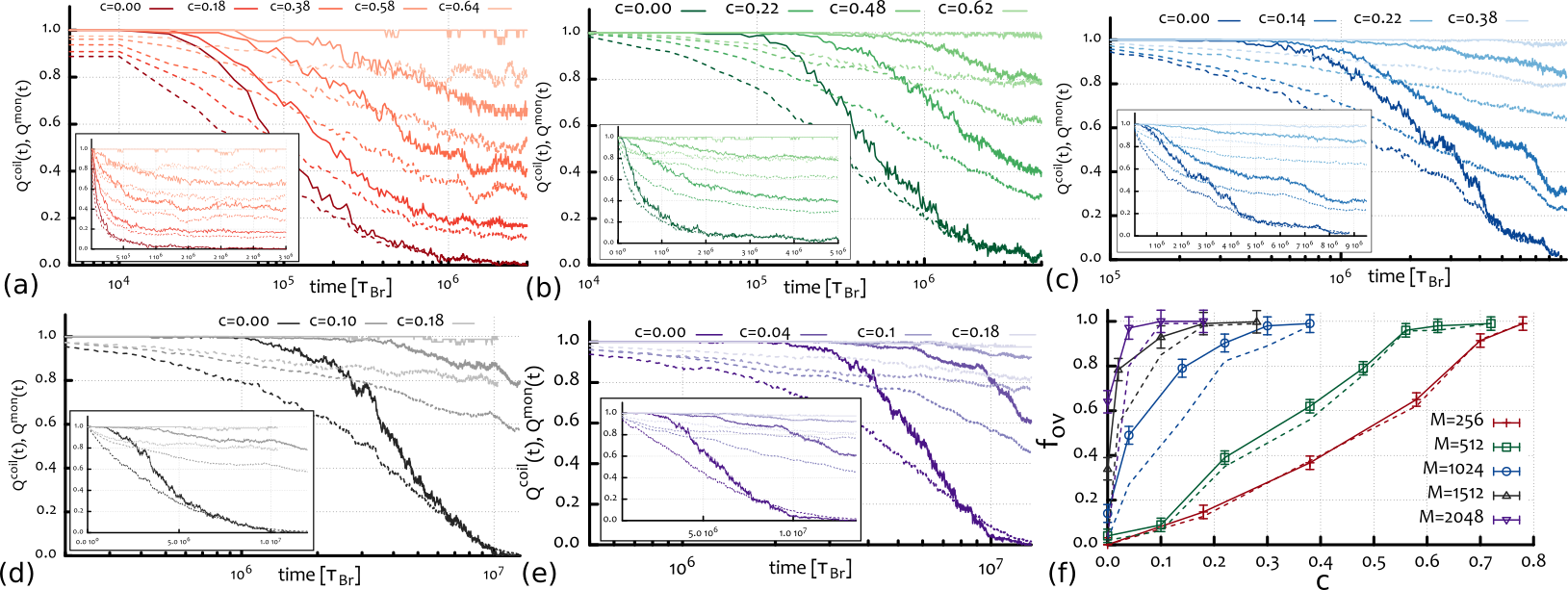}
	\caption{Overlap parameters $Q^{\rm coil}$ (solid lines) and $Q^{\rm mon}$ (dashed lines) for different values of $c$ and for the 5 values of $M$ considered: (a)$M=256$, (b)$M=512$, (c)$M=1024$, (d)$M=1512$ and (e)$M=2048$. In the inset the overlap parameter is plotted in linear scale to highlight the long time flattening. {\bf(f)} The value of the overlap parameter $f_{\rm ov} \equiv Q^{\rm coil}(t=t_i;c)$ is evaluated and reported at two arbitrary (long) times $t_1=5$ $10^6$ $\tau_{Br}$ (solid lines and data points) and $t_2=10^7$ $\tau_{Br}$  (dashed lines), showing a consistent increased tendency to display an arrest of the decay at larger values.}
	\label{fig:SI_Qs}
\end{figure*}

\subsection{The Overlap Parameter Shows an Arrested Decay Corresponding to Caged Length-Scales}
The overlap parameters $Q^{\rm mon}_s(t)$ and $Q^{\rm coil}_s(t)$ are useful to characterise the glassy dynamics~\cite{Karmakar2013}. We here define them here as follows:
\begin{equation}
Q^{\rm mon}(t;c) = \left\langle \Theta(w - |\bm{r}_i(t) - \bm{r}_i(0)|) \right\rangle
\label{eq:OverlapParameter_beads}
\end{equation}
and
\begin{equation}
Q^{\rm coil}(t;c) = \left\langle \Theta(w - |\bm{r}_{CM}(t) - \bm{r}_{CM}(0)|) \right\rangle.
\label{eq:OverlapParameter_CM}
\end{equation}
Where the average is performed over monomers (coils) and initial times. 
The window parameter is chosen to be $w=2R_g$ being the length-scale at which the glassy dynamics is conjectured to occur. In other words, we aim to average out all the jiggling of the coils inside cages of size $2R_g$ and to capture the slowing down of the translational dynamics of the centre of mass of the coils.
As shown in Fig.~S6, this two-point correlation function clearly reflects the arrested relaxation when $c$ is increased toward $c^\dag$. As discussed in the main text, coils that are completely caged cannot escape and freely diffuse. This means that their centre of mass is confined in a cage of linear size 2$R_g$ at all times.   The coils' overlap parameter reflects the constraint by arresting its decay and in particular we find that 
\begin{equation}
\lim_{t\rightarrow \infty} Q^{\rm coil}(t;c^\dag) \simeq 1 
\end{equation}
at any time.

One can also notice that the dynamics of the beads is less constrained than the dynamics of the centre of mass of the coils when $c \rightarrow c^\dag$. As observed, at $c\simeq 0$ one notices that the two correlation functions match in the limit of large $t$. On the contrary, at $c>0$, their difference remains finite at all times and this implies that the relaxation dynamics is decoupled by the topological constraints, which suppress the degree of freedom of the centre of mass of the coils while leaving shorter segments along the chains relatively unhindered. 

Given the fact that the arrested decay of $Q^{\rm coil}(t;c)$ and $Q^{\rm mon}(t;c)$ ends with an unambiguous flattening at a constant value at long times only for small chains, we compare the behaviour of this correlation function by choosing two arbitrary (long) times ($t_1$ and $t_2$)and plotting the value of $f_{\rm ov} \equiv Q^{\rm coil}(t=t_i;c)$ at those times in Fig.~6(f). One can clearly notice that by increasing $c$ any system becomes slower and for larger chains, a small contamination of frozen chains ($c$) is enough to dramatically arrest the decay of the overlap function.

\begin{figure*}[t]
	\centering
	\includegraphics[width=0.95\textwidth]{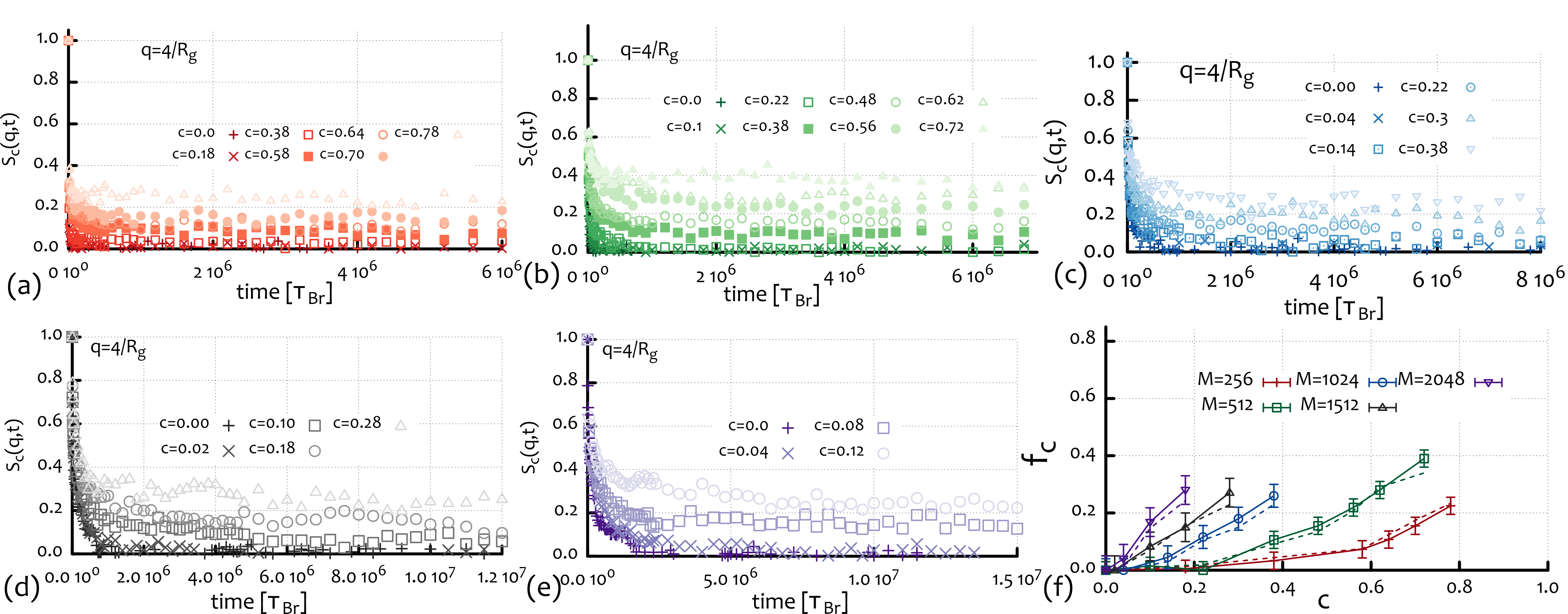}
	\caption{ Scattering function $S_c(q,t)$ computed at $q=4/R_g$ for different chain lengths: (a)$M=256$, (b)$M=512$, (c)$M=1024$, (d)$M=1512$ and (e)$M=2048$. The last three set of curves are plotted in linear scale to highlight the behaviour at large times. \textbf{(f)} The value of the scattering function at arbitrary time $\bar{t}$, $f_c \equiv S_c(q,\bar{t})$ is plotted against $c$ for two chosen values of time $\bar{t}$: $t=5$ $10^6$ $\tau_{Br}$ (solid lines and symbols), $t=10^7$ $\tau_Br$ (dashed lines) and for the different chain lengths. }
	\label{fig:SqDyn_4Rg}
\end{figure*}
%%%%%%%%%%%%%%%%%%
\begin{figure*}[t]
	\centering
	\includegraphics[width=0.95\textwidth]{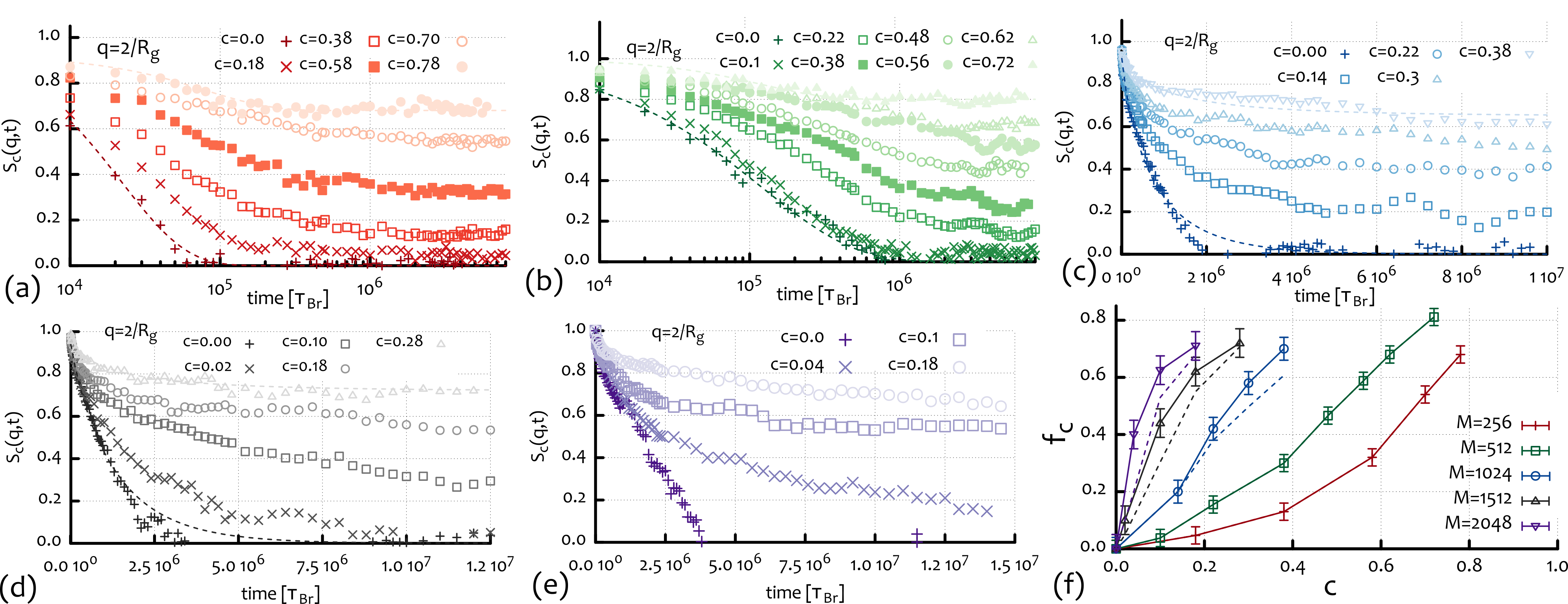}%Figure_SI_DynScattering_2Rg.png
	\caption{Scattering function $S_c(q,t)$ computed at $q=2/R_g$ for different chain lengths: (a)$M=256$, (b)$M=512$, (c)$M=1024$, (d)$M=1512$ and (e)$M=2048$. The last three set of curves are plotted in linear scale to highlight the behaviour at large times. \textbf{(f)} The value of the scattering function at arbitrary time $\bar{t}$, $f_c \equiv S_c(q,\bar{t})$ is plotted against $c$ for two chosen values of time $\bar{t}$: $t_1=5$ $10^6$ $\tau_{Br}$ (solid lines and symbols), $t_2=10^7$ $\tau_Br$ (dashed lines) and for the different chain lengths.  }
	\label{fig:SqDyn_2Rg}
\end{figure*}

\subsection{The Scattering Function Indicates Markedly Different Relaxation Times Above and Below the Average Size of a Coil}

In this section we discuss the dynamic scattering function 
\begin{equation}
S_c(q,t) = \left\langle \dfrac{1}{M} \sum^M_{i,j \in I} e^{i \bm{q}\left[\bm{r}_i(0) - \bm{r}_j(t)\right]} \right\rangle, 
\end{equation}
calculated for $q_1=4/ R_g$ and $q_2=2/ R_g$. The latter explores length scales larger than the diameter of the coils ($l_2=\pi R_g>2R_g$) twice as long of the former, which probes length scales shorter than the diameter of the coils ($l_1=(\pi / 2) R_g<2R_g$).  As one can notice from Fig.~\ref{fig:SqDyn_4Rg} and Fig.~\ref{fig:SqDyn_2Rg}, these two dynamics are markedly different. \\

It is clear from Figs.~S7 and S8 that length scales larger than $2R_g$ are much slower than the internal modes, probed by $S_c(q=4/R_g,t)$. This is shows by the large-time value attained by $S_c(q,t)$, defined as $f_c$, in the two cases and reported in S7(f) and S8(f). In the figures, we show $f_c$ computed for two arbitrarily long times (solid and dashed lines), as done for the overlap function. The way in which $f_c$ grows steeper and steeper for $q=2/R_g$ and for increasing chain lengths as a function of $c$ is indicative that the pinning procedure affects large length scales more severely than shorter ones. 

From these findings, as discussed in the main text, we argue that the relaxation of the long wave length modes is strongly hindered by the pinned rings, while the short wavelength are relatively free to relax. This once again implies that the stronger effect of the pinning of rings is experienced by the translational degrees of freedom of the rings while it leaves short segments of the rings able to partially re-arrange their conformations. \\

\red{
	\subsection{The Efficiency of Freezing Grows Exponentially with the Chains' Length} %Relaxation of the Perturbed Systems Shows a Super-Arrhenius Behaviour}
	The freezing procedure described in the main text offers a pathway to generate glassy states by exploiting the topology of the constituents. We show how the fraction of freely diffusing chains depends on the fraction of (non-)frozen chains in Fig.~S9(a). This is done by tracking the individual MSD of the coils centre of mass and by counting the number of these which have travelled more than 2$R_g$ at the end of the simulation run time and by classifying these as freely diffusing.
	The dashed line represents the curve followed by the data points if every non-explicitly frozen chain were free to diffuse. The deviation from this (zero pinning efficiency) line becomes stronger as the chains become longer and readily show that long chain are very sensitive to a small amount of explicitly frozen chains. \\
	
	Fig.~S9(b) shows the number of caged chains as a function of rings' length. Once again we identify the caged coils by tracking the individual MSD of the centre of mass and by identifying as ``caged'' those which have not travelled more than $2R_g$ at the end of the simulation. We repeat this analysis for every choice of $c$. We observe that not all simulations have the same caging ``efficiency''  but, remarkably, we observe that both the least and the most efficient (highest and lowest number of caged rings per frozen one) scale exponentially with $M$.  
	This finding strongly encourages further computational and experimental studies of this system, as the number of chains implicitly caged can become arbitrarily big depending on the choice of $M$. 
	%As a consequence, we argue that the dramatic slowing down could be observed on a macroscopic scale. In addition,  b
	Because of this exponential increase, fewer explicitly frozen chains will be needed to significantly slow down the system, raising the possibility that the system might spontaneously vitrify. \\
	
	Finally, we study the longest relaxation time of the perturbed systems by computing $\tau_{\rm Relax} \equiv R^2_g/D_{\rm eff}$ and we report the findings in Figs.~S9(c)-(d). The divergence of the relaxation time follows naturally from the fact that $D_{\rm eff}$ is vanishing at $c \rightarrow c^\dag$. In addition, we fitted the values of $\tau_{\rm Relax}$ with an empirical function inspired to the standard Vogel-Fulcher-Tammann function used to describe the relaxation of glass-forming systems 
	\begin{equation}
	\tau_{\rm Relax} = \tau_0 \exp{\left[ \dfrac{D c_0}{c_0-c}\right] }.   
	\end{equation}
	where here $c$ replaces $T$. This result can be understood in terms of cooperativity of the chains: as one gets closer to the critical line $c^\dag(M)$, the activation energy to re-arrange and relax the system becomes higher, as the number of topological constraints becomes closer to the critical value for which all the translational degrees of freedom of the system are quenched. On the other hand, it is important to notice that we can track the relaxation time of the chains only up to roughly two orders of magnitude larger than the unperturbed relaxation time ($\tau_0$) and this is far too small a range to draw definite conclusions on the nature of this divergence.
}

\begin{figure*}[t]
	\centering
	\includegraphics[width=1.0\textwidth]{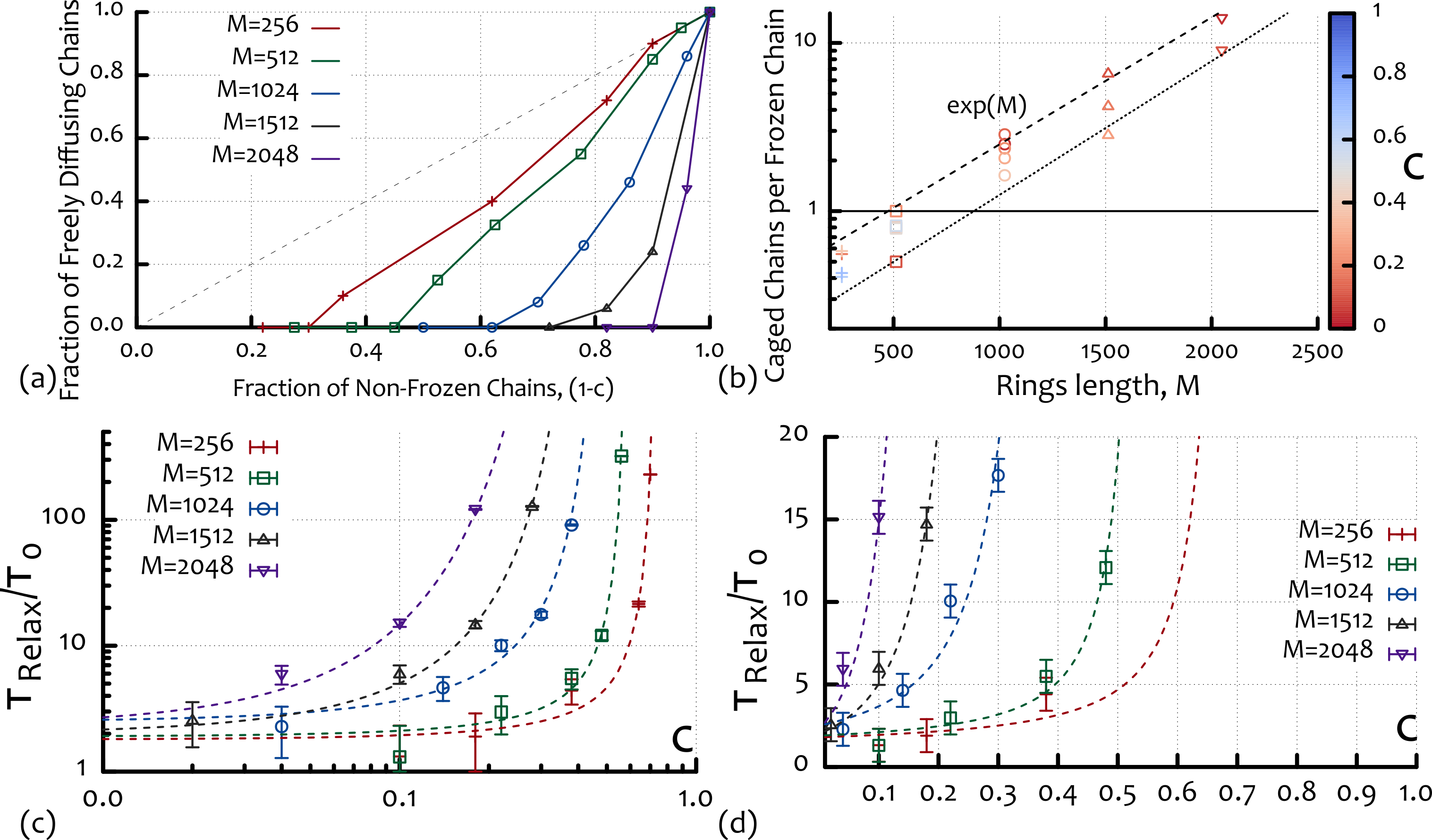}
	\caption{\textbf{(a)}Fraction of freely diffusing rings $\Phi_{fd}=n_{fd}/(1-c)N$ against the fraction of non-frozen chains $1-c$. The dotted line marks the case in which all the non explicitly frozen chains are also freely diffusing. {\bf (b)} The ``efficiency'' of the pinning procedure can be quantified by plotting the number of caged rings per frozen chain as a function of the chains' length and for the various $c$ used. Here the points plotted mark the result from a simulation for fixed $c$ and $M$. Both the most and the least efficient cases (for a given rings length $M$) show an exponential growth with $M$. {\bf (c)-(d)} Relaxation time of the systems computed as $R_g^2/D_{\rm eff}$. The divergence of the relaxation as a function of the freezing fraction $c$ is broadly captured by a VFT function, \emph{i.e.} $\tau_{\rm Relax} = \tau_0 \exp{\left[ A c/(c_0 - c)\right]}$, where $c_0$ is generally larger than $c^\dagger$ defined using $D_{\rm eff}$ (see main text).}
	\label{fig:Suscept}
\end{figure*}

\end{article}

\end{document}